\documentclass[aps,prl,letterpaper,10pt,twocolumn,superscriptaddress]{revtex4-1}
\usepackage[colorlinks=true,allcolors=blue!65]{hyperref}
\usepackage{mathptmx}
\usepackage{amssymb}
\usepackage{amsmath}
\usepackage{graphicx}
\usepackage{subfigure}
\usepackage{amsfonts}
\usepackage{xcolor}
\usepackage{epsfig}
\usepackage{color}
\usepackage{bbm}
\usepackage{bm}
\usepackage{tabularx}
\usepackage{multirow}
\usepackage{mathtools}
\usepackage{xfrac}
\DeclareMathAlphabet{\mathcal}{OMS}{cmsy}{m}{n}

\newcommand{\ket}[1]{\vert{#1}\rangle} 
\newcommand{\bra}[1]{\langle{#1}\vert}

\newcommand{\mean}[1]{\langle #1 \rangle}

\newcommand{\hatd}[1]{\hat{#1}^{\dagger}}

\newcommand{\half}{\frac{1}{2}}

\newcommand{\sighat}{\hat{\sigma}}

\newcommand{\gele}{\textsl{g}_{\rm e}}

\begin{document}
\title{Analog Quantum Simulation of Extremely Sub-Ohmic Spin-Boson Models}

\author{Mehdi Abdi}
\email{mehabdi@gmail.com}
\affiliation{Department of Physics, Isfahan University of Technology, Isfahan 84156-83111, Iran}

\author{Martin B. Plenio}
\affiliation{Institute of Theoretical Physics and IQST, Albert-Einstein-Allee 11, Ulm University, 89069 Ulm, Germany}

\begin{abstract}
We propose a scheme for the quantum simulation of sub-Ohmic spin--boson models by color centers in free-standing hexagonal boron nitride (h-BN) membranes. The electronic spin of a color center that couples to the membrane vibrational spectrum constitute the physical model. The spin-motion coupling is provided by an external magnetic field gradient.
In this study, we show that a class of spectral densities can be attained by engineering geometry and boundary conditions of the h-BN resonator. We then put our focus on two extreme cases, i.e. $1/f$- and white-noise spectral densities. Spin coherence and polarization dynamics are studied. Our calculations show coherence revivals at periods set by the bath characteristic frequency signaling the non-Markovian nature of the baths. The nonequilibrium dynamics of the spin polarization exhibits a coherent localization, a property peculiar to the quantum phase transition in extremely sub-Ohmic spin-boson models.
Our scheme may find application in understanding sources of decoherence in solid-state quantum bits.
\end{abstract}

\maketitle

%
%
The dynamics of a two-level system interacting with a bath of bosonic modes, the so-called spin--boson model, has been used for a long time to explain the observed decoherence of qubits~\cite{Caldeira1981, Leggett1987}. The model was first proposed to explain the decoherence of electronic states in molecular systems, however, nowadays is used to estimate the decoherence in almost every quantum system. In the general case, the model is not analytically solvable and its full understanding demands either numerical methods and where these are not efficient a quantum simulation~\cite{Georgescu2014}. Retrieval of the macroscopic nature of a reservoir in such simulators necessitates involvement of a large number of boson modes. It also should have the versatility of handling various coupling forms and strengths as well as spectral densities $\mathcal{J}(\nu)$---the function that encodes information about the bath and fully determines the dynamics in the case of thermal initial state. Among the important noise spectra are $\mathcal{J}(\nu)\propto\nu^s$, e.g. $1/f$-noise ($s=-1$)~\cite{Paladino2002, Koch2007}, white-noise ($s=0$)~\cite{Martinis2003}, and Ohmic ($s=+1$) spectral densities from which the last one and its varieties, i.e. sub- and super-Ohmic cases, has been broadly studied in the literature~\cite{Sassetti1990, Kehrein1996, Vojta2005}. Nonetheless, despite their crucial importance in decoherence of solid-state qubits the quantum analysis of the $1/f$- and white-noise is rather overlooked, with the exception of Refs.~\cite{Paladino2002, Paladino2014}. In particular, quantum dynamics of a system with $1/f$-noise cannot be simulated efficiently by the developed full quantum mechanical techniques~\cite{Makri1995, Wang2001, Chin2010, Woods2014} and a direct quantum simulation must be invoked. Moreover, the occurrence of quantum phase transition in the sub-Ohmic regime and its correspondence to the classical phase transition make these two cases even more interesting from the physics perspective~\cite{Anders2007, Alvermann2009, Winter2009, Chin2011}.

The spin--boson model quantum simulation schemes have been proposed and partly realized in ion-traps and superconducting circuits~\cite{Porras2008, Lemmer2018, Leppaekangas2018}. In ion-trap proposals either a chain of trapped ions is used to simulate the model~\cite{Porras2008} or more advanced techniques are employed to perform the simulation by a handful of ions~\cite{Lemmer2018}.
In any case, the number of available oscillators is restricted and this limits the spectral densities that can be decomposed.
The problem persists in superconducting circuit schemes owing to the finite number of available cavity modes upper bounded by superconducting energy gap. Moreover, the Josephson qubits in such setups are naturally subject to the $1/f$- and white-noise~\cite{Paladino2002, Paladino2014} and thus are not appropriate for the quantum simulation of these noise spectra.

To circumvent these limitations, here we propose a versatile setup based on color centers in hexagonal boron nitride (h-BN) membranes~\cite{Tran2015, Abdi2018}. The spin--boson model is realized by coupling the electronic spin of a color center to vibrations of a free-standing h-BN membrane~\cite{Abdi2017}. The scheme offers controllable coupling strength between the two-level-system and the bath covering spectral densities of $\mathcal{J}(\nu)\propto\nu^s$ with $-1\leq s\leq 0$. In this paper we thoroughly investigate the $s=-1$ and $s=0$ cases. Non-Markovian nature of the studied noise spectra is revealed by observation of collapse and revivals in the spin coherence---off-diagonal elements of the spin density matrix. Unlike setups proposed before, the number of interacting modes in our scheme is far beyond the mesoscopic scale and thus the limitations therein are avoided.
The fundamental mode, the mode with lowest frequency, of the membrane places a cutoff frequency from below, which dominantly sets the period of bath memory revivals. Such revivals signal the non-Markovian nature of the bath that is then quantified by the measure introduced in Refs.~\cite{Breuer2009, Rivas2010, Rivas2014}.
The nonequilibrium dynamics of the spin polarization reveals that both bath spectral densities force the spin into a coherent localization, a feature of extremely sub-Ohmic spin-boson models~\cite{Kast2013}. Our investigation on the polarization localization then verifies occurrence of a quantum phase transition consistent with universality of the sub-Ohmic spectral densities~\cite{Winter2009}.

%
%
\textit{Model.---}%
The system is composed of a free-standing h-BN membrane with an embedded color center. Some species of the h-BN color centers exhibit a spin doublet electronic ground state~\cite{Abdi2018}. When immersed in a magnetic field gradient perpendicular to the plane, the spin degree of freedom couples to the position degree of freedom of the membrane.
The Hamiltonian of such system is~\cite{Abdi2017, Abdi2018a}
\begin{subequations}
\begin{align}
\hat H &= \hat H_{\rm m} +\hat H_{\rm q} +\gele\mu_{\rm B}\eta\sighat_z \hat X, \\
\hat H_{\rm q} &= \half(\Delta\sighat_z +\Omega\sighat_x),
\end{align}
\label{totham}
\end{subequations}
where $\hat H_{\rm q}$ is Hamiltonian of the spin-qubit driven at Rabi frequency $\Omega$ with frequency detuning $\Delta$ expressed in terms of Pauli matrices. $\hat H_{\rm m}$ is the mechanical Hamiltonian, $\hat X$ denotes deviation of the membrane from its equilibrium position, and $\eta$ is the magnetic field gradient. Finally, $\mu_{\rm B}$ and $\gele$ are the electron Bohr magneton and g-factor, respectively.
In terms of normal mechanical modes, the spin interacts with a set of vibrational modes that undergo harmonic oscillations. Hence, the system exhibits a `natural' spin--boson model~\cite{Breuer2007}.
The fundamental mode in the mechanical spectrum is defined as the mode with lowest frequency $\omega_0 \leq \{\omega_i\}$, while the mode with highest frequency $\omega_N$ is set by the vibrational wavelength, which in turn is lower-bounded by the lattice constant. These two extreme frequencies are engineered by the geometry and boundary conditions of the membrane.
We expand the displacement operator in normal modes of the membrane $\hat X=\sum_{k=0}^N x_{\rm zp}^{k}(\hat b_k +\hatd b_k)$ to arrive at the mechanical Hamiltonian $\hat H_{\rm m} = \sum_{k=0}^{N}\omega_k\hatd b_k\hat b_k$ and the interaction:
\begin{equation}
\hat H_{\rm int} = \half\sum_{k=0}^{N}g_k\sighat_z (b_k +\hatd b_k).	
\end{equation}
Here, the bosonic annihilation (creation) operator $\hat b_k$ ($\hatd b_k$) is assigned to the mechanical normal mode whose frequency is $\omega_k$. These modes are also subject to a damping with the rate $\gamma_k$. In our analysis we assume all of the modes having the same quality factor $Q$ and thus $\gamma_k=\omega_k/Q$. Degenerate modes can be easily distinguished by one extra index. Nonetheless, without loss of generality we restrict ourselves to the non-degenerate systems. The coupling strength of each mode to the spin-qubit is given by $g_k \equiv \gele\mu_{\rm B}\eta x_{\rm zp}^{k}$ with the zero-point fluctuation amplitude of $k$th mode $x_{\rm zp}^{k}=\psi_k(\mathbf{r}_0)\cdot\sqrt{\hbar/2m_k^*\omega_k}$. Here, $\psi_k(\mathbf{r})$ and $\mathbf{r}_0$ are profile of the mechanical mode and position of the defect, respectively. With membrane mass density $\mu$ the effective mass of each mode is $m_k^* = \mu\int\!d^2\mathbf{r}\psi_k(\mathbf{r})^2$. The mode profiles are normalized such that $\max[\psi_k(\mathbf{r})]=1$.
The spectral density of bath is then given by
\begin{equation}
\mathcal{J}(\nu)=\pi\sum_k g_k^2 L_k(\nu),~~~~~ L_k(\nu)\equiv\frac{\gamma_k/2}{(\gamma_k/2)^2+(\nu -\omega_k)^2}.
\end{equation}
Here, $L_k(\nu)$ is a Lorentzian function referring to the spectrum of the $k$th mechanical mode.

\begin{figure}[tb]
\includegraphics[width=0.85\columnwidth]{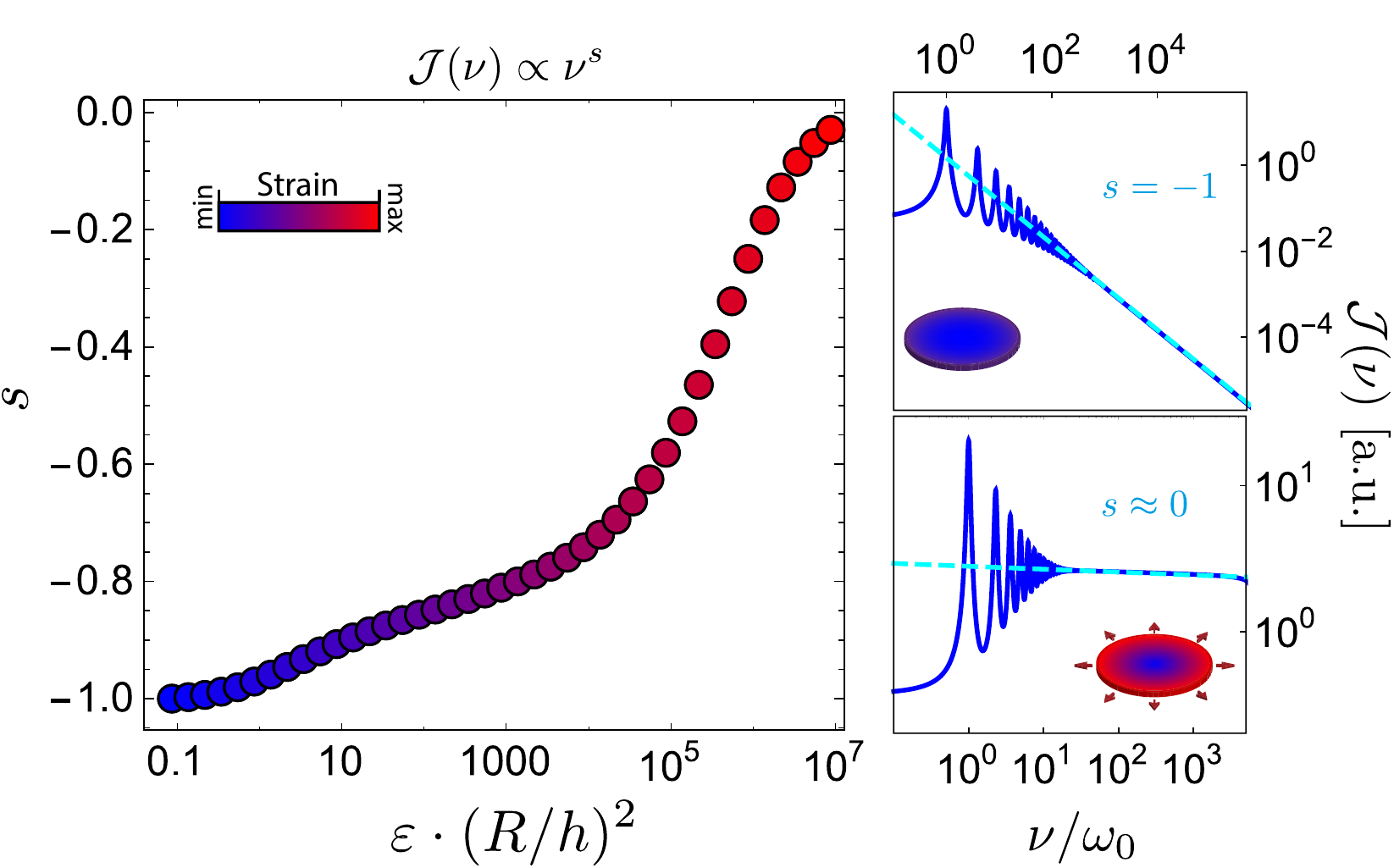}
\caption{%
Behavior of the spectral density exponent for circularly clamped membrane with respect to the circumferential tensile strain. Spectral densities of the two extremes that give $1/f$-noise and white-noise and happen respectively for $\varepsilon \ll\!(h/R)^2$ (simply clamped) and $\varepsilon \gg\!(h/R)^2$ (under dominant tensile strain) are shown in the right panels.}
\label{fig:spec}%
\end{figure}
%

%
%
\textit{Scheme.---}%
In this work, we study a circular geometry: A h-BN membrane of radius $R$ and thickness $h$ with the defect at the center of membrane. In this case, the spin only couples to the axisymmetric modes. The theory, however, applies to arbitrary membrane shapes and defect positions. By tuning boundary conditions of the membrane, spectral densities of $\mathcal{J}(\nu)\propto \nu^s$ with $-1\leq s \leq 0$ can be engineered. This indeed can essentially be done by tuning the applied homogenous tensile strain to the boundaries (see supplemental information for detailed analysis~\cite{suppinfo}). The strain engineering has already proven to be successful in SiN membrane mechanical resonators~\cite{Ghadimi2017}, a method that may be applicable to the h-BN materials as well.
In Fig.~\ref{fig:spec} variations of the exponent $s$ with the strain $\varepsilon$ is plotted. The focus in this work is put on two extremes, namely $s=-1$, which is attained for a softly clamped membrane $\varepsilon \approx 0$ and $s=0$ that can be obtained for a dominant tensile force at the boundaries $\varepsilon \gg (h/R)^2$.
The maximum number of modes involved in the spin dynamics is imposed by the lattice constant $a$ as $N = \lfloor R/a \rfloor$ where $\lfloor x \rfloor$ yields the integer part of $x$. For instance, a $R=150$~nm h-BN membrane provides $N \approx 1000$ modes, which is in the macroscopic realm.

The normal mechanical modes of a softly clamped circular membrane have frequencies $\omega_k = (\alpha_k/R)^2\sqrt{Eh^3/12\mu(1-\sigma^2)}$ with elasticity modulus $E$ and Poisson ratio $\sigma$. Here, $\alpha_k$s are solutions to the equation $J_0(\alpha_k)I_1(\alpha_k)+J_1(\alpha_k)I_0(\alpha_k)=0$, where $J_n$ and $I_n$ are $n$th order Bessel functions of the first kind. Mode profiles are then $\psi_k(r) = J_0(\alpha_k\frac{r}{R})-\frac{J_0(\alpha_k)}{I_0(\alpha_k)}I_0(\alpha_k\frac{r}{R})$ giving coupling rates $g_k=g_0/\sqrt[4]{\omega_k/\omega_0}$~\cite{Abdi2016}. These result-in a $1/f$-noise spectral density $\mathcal{J}(\nu)\propto\nu^{-1}$ for operating frequencies much higher than the low-frequency cutoff, the fundamental mode frequency $\omega_0$, as shown in Fig.~\ref{fig:spec}. We will refer to this as case-I.
Instead, when the circular membrane is under homogenous dominant tensile strain around its circumference the mode profiles take the simple form of $\psi_k(r)=J_0(\beta_k\frac{r}{R})$, where $\beta_k$s are zeros of the Bessel function of the first kind $J_0(\beta_k)=0$~\cite{Abdi2016}. The frequency spectrum in this case is $\omega_k = (\beta_k/R)\sqrt{Eh\varepsilon/\mu}$, while the coupling rate for all of the modes is equal to that of the fundamental mode $g_k =g_0$. Hence, a white-noise spectral density is attained $\mathcal{J}(\nu)\propto\nu^0$ (shall be referred to as case-II).
In comparison with the standard form of the spectral function $\mathcal{J}(\nu)=2\pi\alpha\nu_c^{1-s}\nu^s$ we find that $2\pi\alpha\nu_c^{1-s} = g_0^2$ in both cases, where the cut-off frequency is given by $\nu_c \simeq \omega_N$, set by the maximum possible nodes in the membrane. We find that $\omega_{1000}/\omega_0$ is about $\sim\!10^6$ and $\sim\!10^3$ for case-I and II, respectively [Fig.~\ref{fig:spec}].

\begin{figure}[tb]
\includegraphics[width=\columnwidth]{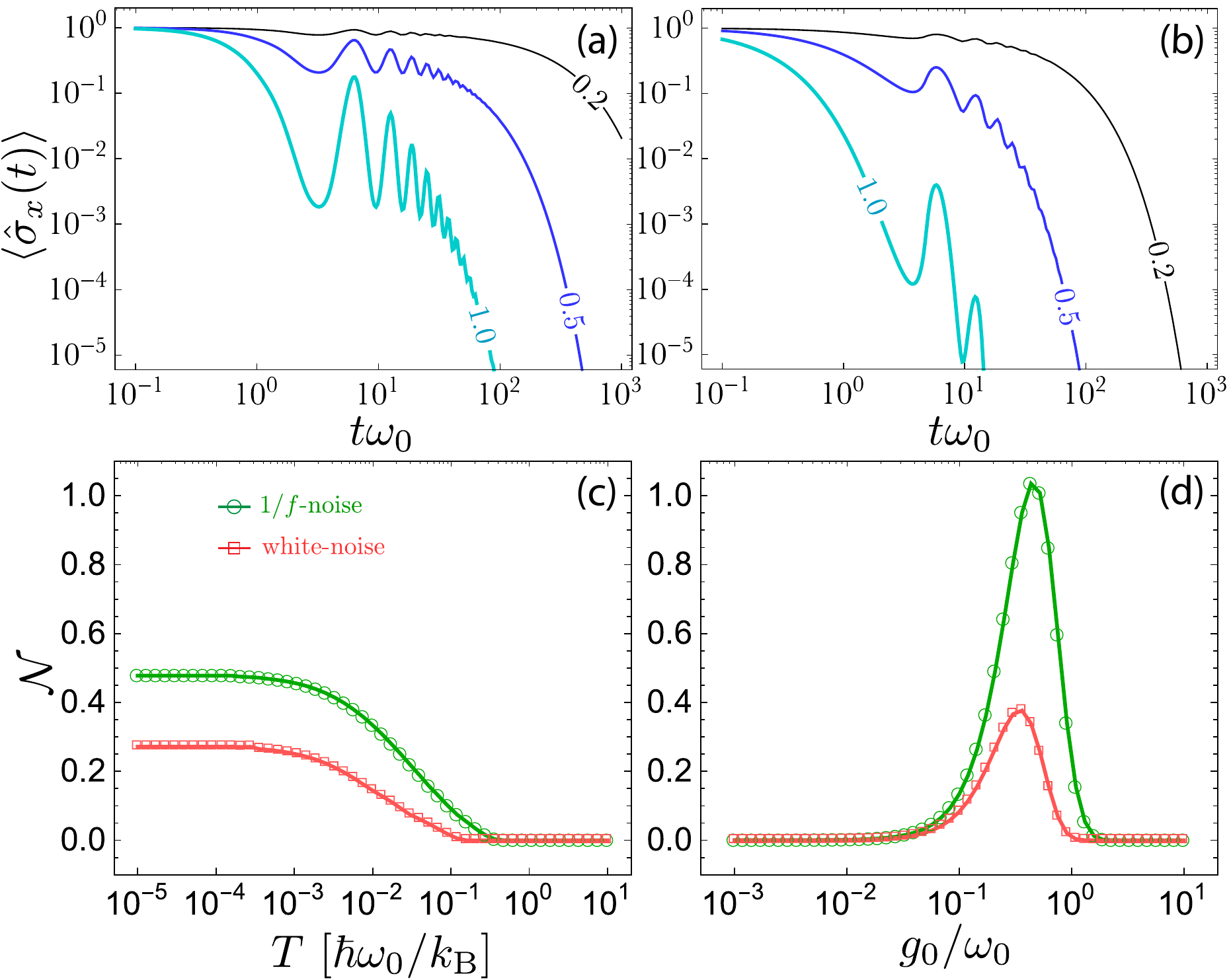}
\caption{%
Upper panels: Collapse and revival in the spin coherence for case-I (a) and case-II (b). $\mean{\sighat_x(t)}$ is plotted at three different coupling rates (in units of $\omega_0$) at zero bath temperature.
Lower panels: Non-Markovianity measure variations in pure dephasing spin dynamics for case-I (green) and case-II (red) as a function of: (c) temperature at fixed coupling rate $g_0=0.2\omega_0$ and (d) coupling rate at fixed bath temperature $T = 10^{-3} \cdot (\hbar\omega_0/k_{\rm B})$.}
\label{fig:nonmarkov}%
\end{figure}

%
%
\textit{Pure dephasing.---}
Let us first consider the pure dephasing case, which is exactly solvable for the bilinear model realized here. We thus assume that $\Omega = 0$ and an initially separable qubit--bath state. The off-diagonal elements of the spin density matrix $\rho$ at every instance of time are then found to evolve as
$\bra{i}\rho(t)\ket{j} = \bra{j}\rho(t)\ket{i} = \bra{i}\rho(0)\ket{j}e^{\widetilde\Gamma(t)}$
with
\begin{equation}
\widetilde\Gamma(t)=-\int_{0}^\infty\hspace{-1mm} d\nu \mathcal{J}(\nu)\coth\!\big(\frac{\hbar\nu}{2k_{\rm B} T}\big)\big(\frac{1-\cos\nu t}{\nu^2}\big),
\label{deph}
\end{equation}
for a thermal bosonic bath at temperature $T$, while $k_{\rm B}$ is the Boltzmann constant~\cite{Breuer2007, suppinfo}.
In Fig.~\ref{fig:nonmarkov}(a) and (b) we plot the time evolution of the spin coherence, $\mean{\sighat_x(t)}$, for the two cases explained above at zero temperature.
The most interesting features are the coherence revivals visible in both cases.
The revivals occur around full period of the fundamental mode $t\simeq 2k\pi/\omega_0$ with $(k=1,2,\cdots)$ and live longer for case-I. This traces back to the fact that the most pronounced peak in both spectral density occurs at $\omega_0$ and that the frequency spacing of the modes is proportional to $\omega_0$: $\delta_k\equiv \omega_k-\omega_{k-1} \cong 2k\omega_0$ and $\delta_k \cong \omega_0$ in case-I and -II, respectively. This behavior survives even at finite temperatures~\cite{suppinfo}. Such revivals are a clear signature of bath non-Markovianity, the property we study next.


%

To quantify non-Markovian behavior of the qubit state evolving under interaction with the bosonic bath we use the measure introduced in Ref.~\cite{Breuer2009}.
For the pure dephasing case this measure of non-Markovianity assumes a simple form, which is given by 
$\mathcal{N}=\sum_k'\big[\mathcal{G}(t_k^f)-\mathcal{G}(t_k^i)\big]$
where $\mathcal{G}(t)\equiv e^{\widetilde\Gamma(t)}$ is the function of coherence envelope, and $t_k^i$ and $t_k^f$ respectively denote initial and final points of the $k$th discretized time interval. The prime on the sum denotes its extension only over the time intervals in which $\mathcal{G}(t)$ is ascending.
The above quantity is computed as a function of coupling rate and temperature and plotted in Fig.~\ref{fig:nonmarkov}.
At low temperatures both cases present a Markovian behavior $\mathcal{N}\neq 1$ only if the coupling rate $g_0$ assumes intermediate values $0.1 \lesssim g_0/\omega_0 \lesssim 1$. This proves that in the weak coupling regime ($g_0 \ll \omega_0$) the Born-Markov approximation holds for both cases even at low temperatures. However, there is a threshold bath temperature, dependent on the coupling rate, that the system turns into a Markovian bath [Fig.~\ref{fig:nonmarkov}(a)]. Moreover, for any coupling rate and temperature values a $1/f$-noise bath exhibits a higher non-Markovianity than a white-noise bath.

%
%
\textit{Relaxation.---}%
Now we study a different regime where in the qubit Hamiltonian Eq.~(\ref{totham}b) $\Delta =0$ and $\Omega\neq 0$. This problem might be analytically solvable even though the interaction and qubit Hamiltonians do not commute~\cite{Chen2012}. However, the tedious calculations turns one to the approximate and numerical methods. The Born-Markov approximation that are typically used, rely on the weak coupling and memory-less bath assumptions, which are neither general nor reveal the non-Markovian root of the models studied here.
Instead, numerically approximate methods, e.g. non-interacting blip approximation (NIBA) or Ehrenfest (mean-field) method are basically operational~\cite{Leggett1984, Stock1995}. The former is proven reliable for fast baths ($\nu_c \gg \Omega$), while the latter explains well a bath in adiabatic limit ($\nu_c \ll \Omega$).
To tackle the problem, we employ a hybrid Ehrenfest--NIBA method introduced in Ref.~\cite{Berkelbach2012}. In this method dynamics of the low frequency bath modes (with respect to a characteristic frequency which is expected to be around $\Omega$) and the spin back-action is dealt with the Ehrenfest method and the effect of high frequency modes is taken into account by NIBA (see the supplemental material for the details~\cite{suppinfo}). This technique has proven to be reliable for adiabatic, diabatic, and the intermediate limits~\cite{Berkelbach2012}.

\begin{figure}[tb]
\includegraphics[width=\columnwidth]{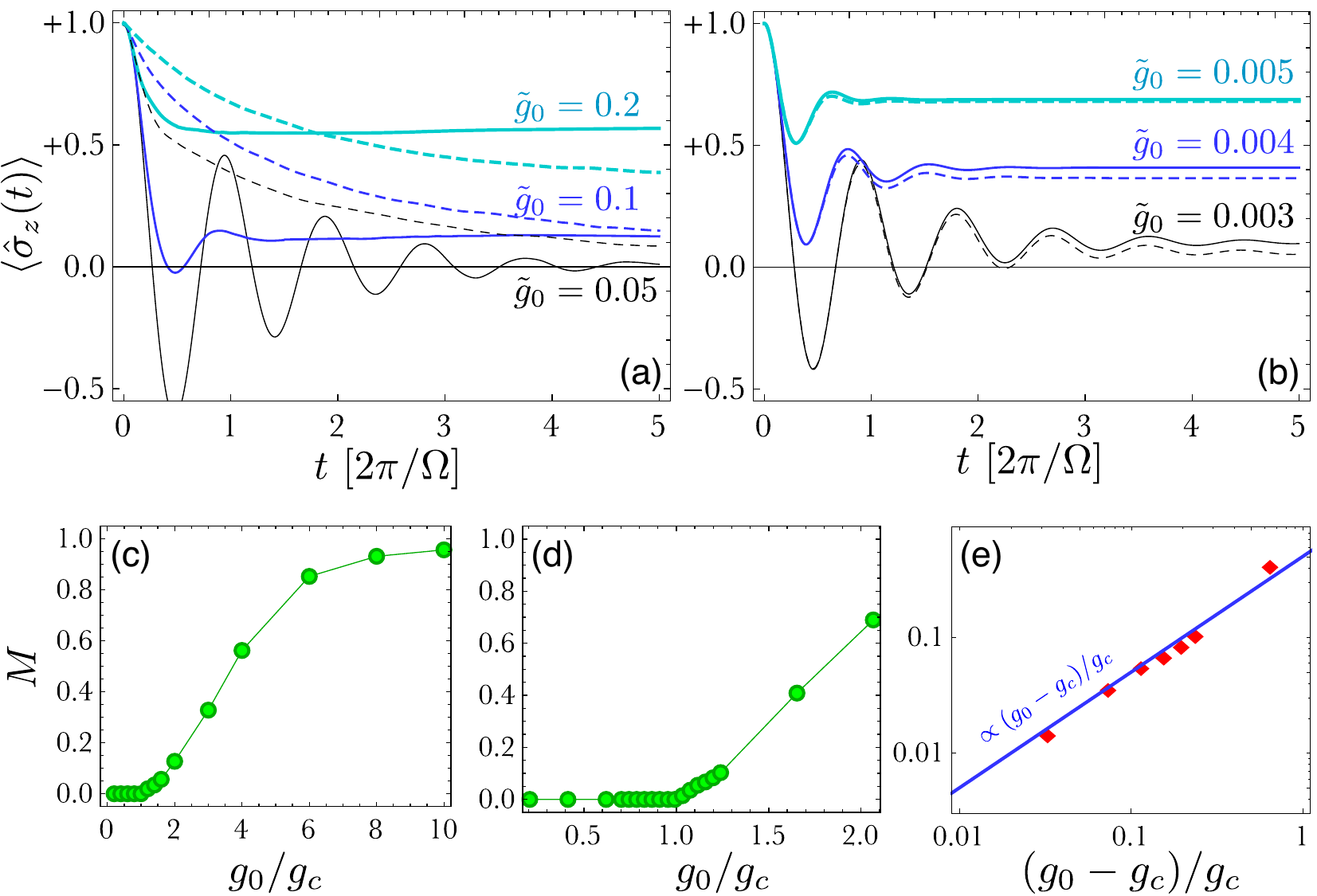}
\caption{%
Spin polarization dynamics $\mean{\sighat_z(t)}$ for various coupling rates $\tilde g_0 \equiv g_0/\omega_N$ at zero temperature (solid lines) and $T = 0.002\cdot(\hbar\omega_N/k_{\rm B})$: (a) case-I and (b) case-II. In (c) and (d) transition of magnetization from a delocalized to the localized phase is shown for case-I and -II, respectively. The plot in (e) reveals power law behavior of the magnetization close to the quantum critical point. The blue line is a linear fit to the data. In all plots we have set $\Omega=0.1\omega_N$.}
\label{fig:relax}%
\end{figure}

The results are presented in Fig.~\ref{fig:relax}(a) and (b) where we plot the spin polarization dynamics for case-I and -II at zero and finite temperatures when $\Omega=0.1\omega_N$.
Obviously, the macroscopicity of our scheme precludes appearance of non-Markovian effects like polarization revivals that are visible for a mesoscopic chain of trapped ions even at short-time dynamics~\cite{Porras2008, Lemmer2018}. The reason is twofold; first since $\Omega \gg \omega_0, \{\delta_k\}$ (and still $\Omega \ll \nu_c$) the revivals will only be expectable at very long-time dynamics. Second, the overlap of the finite width Lorentzian functions $L_k(\nu)$ flattens the spectral density at such high frequencies ($\Omega \gg \omega_0$) and washes out any signature of the bath discreteness and finite dimensionality [see Fig.~\ref{fig:spec}(c)].
The spin polarization get localized for sufficiently large coupling rates in both cases. As expected from extremely sub-Ohmic spectral densities there is no incoherent decay in either case at zero temperature.
Such coherent localizations are exclusive to the strongly sub-Ohmic baths where $s \ll 1$ and is believed to be due to the presence of a nontrivial quantum phase transition~\cite{Anders2007, Kast2013}.
At finite temperatures this behavior dramatically changes for the $1/f$-noise where the bath forces the spin into an incoherent localization despite some shoulder-like oscillations at small coupling rates, see Fig.~\ref{fig:relax}(a). In contrast, a thermal white-noise bath does not demolish this coherent behavior.

The spin-polarization dynamics exhibit both delocalized and localized phases.
Localization of the magnetization $M\equiv\mean{\sighat_z(t\rightarrow\infty)}$, as an order parameter, confirms the quantum phase transition in our setup. Since the power exponent of the spectral densities considered here are below zero one would expect to have the quantum critical point at zero coupling rate, and thus, only the localized phase. However, the system imposes a lower cutoff frequency in our scheme and this guarantees existence of the delocalized phase. Therefore, the transition happens at a finite coupling rate $g_c$.
We identify the quantum critical point by inspecting the ground state energy $E_G(g_0)$ that has been computed by the generalized multipolaron expansion~\cite{Bera2014}. The ground state energy of the system is a continuous function of $g_0$ in both cases and manifests a discontinuity only in its second derivative at the critical point $[d^2E_G/dg_0^2]_{g_0=g_c}$. For case-II we find $g_c \approx 0.00242\omega_N$ while for case-I the task is not trivial and a rough estimation reveals that the critical point is around $g_c \simeq 0.05\omega_N$.
A thorough study of the quantum phase transition and its signatures will be studied elsewhere. Specifically, the quantum phase transition of the $1/f$-noise system is complicated and requires a careful analysis.
Nevertheless, in Figs.~{\ref{fig:relax}(c)-(e) we present our immediate observations: (i) At the critical point the spin magnetization $M$ sets off from zero to finite values [Figs.~\ref{fig:relax}(c) and (d)]. (ii) Close to the critical point $g_0\rightarrow g_c^+$ a spin coupled to the bath with white-noise spectral density follows the universal quantum critical behavior. That is, the magnetization grows linearly with distance from the critical point $M\propto[(g_0-g_c)/g_c]^1$. Since $g_0^2 \propto \alpha$ our finding is consistent with the critical behavior of sub-Ohmic spin-boson models studied before where the critical scaling exponent for the magnetization is found to be $\half$, i.e. $M^2 \propto (\alpha - \alpha_c)/\alpha_c$~\cite{Winter2009}.

%
%
\textit{Outlook and summary.---}%
In this work we proposed a scheme based on h-BN membranes for quantum simulation of a spin interacting with a bath of harmonic oscillators, the spin--boson model. The scheme allows for quantum simulation of baths with spectral densities $\mathcal{J}(\nu)\propto\nu^s$ with $-1\leq s \leq 0$. We have studied two specific cases for a circular membrane: softly clamped ($s=-1$) and highly strained at the edge ($s=0$) and showed that dynamics of these systems provides interesting features. In particular, the collapse and revivals in the spin coherence at time periods set by the bosonic bath fundamental frequency. Also, coherent localization of the spin polarization, which is characteristic of strongly sub-Ohmic spin-boson models. We also confirmed universal behavior of the magnetization for the system with white-noise spectral density.
The scheme can be generalized to other geometries and boundary conditions for the membrane. For example, having the edges of a square membrane in two different boundary conditions will result-in local variations in the spectral density. Hence, electronic spin of defects in different positions will be subject to different decoherence schemes. When brought into interaction with each other the entanglement of two spins is anticipated to show interesting dynamics~\cite{Bellomo2007,Sinayskiy2009}.
Furthermore, close to the fundamental mode frequency $\omega_0$ the engineered spectral density function assumes a structured shape formed by addition of Lorentzians [see Fig.~\ref{fig:spec}]. For a non-driven spin-boson system, $\Delta=0$ in Eq.~(\ref{totham}), that the spin-polarization dynamics depends on the oscillation frequency $\Omega$ the system can be tuned such that $\Omega \sim \omega_0$, and thus, the scheme may become applicable to complex systems with structured noise spectra that specially appear in biomolecular systems~\cite{Huelga2013}.

%
%
\begin{acknowledgements}
This work was supported by the ERC Synergy grant BioQ. The authors acknowledge support by the state of Baden-W{\"u}rttemberg through bwHPC.
M.A. thanks A. Smirne for fruitful discussions.
\end{acknowledgements}

%
%
\bibliography{spinboson}

\end{document}